\documentstyle[twoside,fleqn,espcrc2,psfig]{article}

\def\etal{{\sl et al.}}
\def\ibid{{\sl ibid.}}
\title{Experimental Measurements of Atmospheric Neutrinos}
\author{Edward Kearns\address{Department of Physics, Boston University\\
                              590 Commonwealth Ave.\\
                              Boston, MA 02215}
\thanks{Presented at TAUP97, the 5th Int'l Workshop on Topics
in Astroparticle and Underground Physics, Sept.7-11, 1997, Laboratori
Nazionali del Gran Sasso, Assergi, Italy}
}

\begin{document}

\begin{abstract}
  This talk reports the latest indications of an anomaly in the
  measurements of atmospheric neutrinos. New results from Soudan-2 and
  Super-Kamiokande provide evidence that the ratio of $\nu_\mu$ to
  $\nu_e$ interactions is not as expected. High energy
  Super-Kamiokande data indicates the cause is a deficit of
  upward-going $\nu_\mu$, and the zenith angle dependence of the
  effect is consistent with neutrino oscillations. Upward-going muon
  measurements by several detectors are discussed, but in total they
  provide inconclusive evidence for the anomaly.
\end{abstract}

\maketitle

\section{Introduction}

Large underground detectors originally built to search for proton
decay are also exposed to a flux of neutrinos created by cosmic ray
showers in the upper atmosphere.  Neutrino interactions in the detector
can mimic proton decay, and therefore a great effort has gone into
predicting the rate and topology of the neutrino background.  A
byproduct of this effort has been the recognition of an anomaly: the
relative rate of $\nu_\mu$ and $\nu_e$ interactions disagrees with
expectation, and the baseline and energy dependence of the
disagreement suggests that neutrino oscillation may be the culprit.

The nature of these experiments is to measure the rate of neutrino
interactions in the detector and to compare that rate with theoretical
prediction.  The theoretical task is to predict the neutrino flux as a
function of energy, direction, and flavor, taking into account
measurements of cosmic ray flux, geomagnetic cutoff, and production
and decay of secondary mesons.  Detailed Monte Carlo programs are used
to estimate the cross section for neutrino interaction and simulate
the response of the detector.  The experimental task is to measure as
much as possible about the neutrino interaction, in particular, the
energy, direction, and flavor of the final state lepton, from which
the neutrino properties are inferred.

\section{Flux prediction}

One of the tenets of the atmospheric neutrino anomaly is that the flux
ratio\footnote{technically $(\nu_\mu +
  \bar{\nu}_\mu)/(\nu_e+\bar{\nu}_e)$} ($\nu_\mu/\nu_e$) is more
accurately predicted that the $\nu_\mu$ or $\nu_e$ flux alone. The
principal effect is that cosmic ray showers consist mostly of pions,
which decay to $\mu+\nu_\mu$, and the $\mu$ decays to to
$e+\nu_e+\nu_\mu$, resulting in a flux ratio ($\nu_\mu/\nu_e) \sim 2$.
The authors of detailed calculations of the flux models have
collaborated to compare results\cite{Gais96-compare} and reached an
understanding of many of the differences between their earlier
publications. Two updated
calculations\cite{Hond95-flux,Barr89-bartolflux} cover a wide energy
range (10 MeV--10 TeV) and are used by current experiments.  There are
now also an assortment of new high altitude $\mu$ measurements
available for comparison
\cite{Circ97-durban-MASS,Barb97-durban-CAPRICE,Tarl97-durban-HEAT}.
Further details of flux calculations were presented at this meeting by
T. Stanev\cite{Stan97-taup}. However, there is currently no indication
that poor knowledge of the predicted flux could be responsible for the
experimental anomalies described below.

\section{Summary of results}

%%etk For the purpose of studying 
To study the ($\nu_\mu/\nu_e$) flux ratio, most
experiments calculate the double-ratio or ratio-of-ratios:
\begin{equation}
R = \frac{(N_\mu/N_e)_{DATA}}{(N_\mu/N_e)_{MC}},
\end{equation}
where $N$ refers to the number of events where the final state lepton
is classified as $\mu$-like or $e$-like by some identification
algorithm. As mentioned above, the ratio of $\nu_\mu$ to $\nu_e$ flux
is accurately predicted; in addition, other theoretical and
experimental uncertainties largely cancel. Table~\ref{tab:r-summary}
lists previous measurements\footnote{The first uncertainty quoted is
  statistical, the second is systematic; this convention will be used
  throughout the paper.}  of $R$ for $E_\nu~\sim1$~GeV, including the
new results from Soudan and Super-Kamiokande discussed in this paper.
The kinematic limits differ somewhat from experiment to experiment,
with minimum lepton momenta requirements from 100 to 200 MeV/c.
Kamiokande restricted their sample to\footnote{Visible energy
  ($E_{vis}$) is defined as the energy of an electromagnetic shower
  that produces a given amount of Cherenkov light.} $E_{vis} < 1.33$
GeV, IMB to $p < 1.5$ GeV; the other experiments did not specify an
upper limit, but all results are dominated by $\sim1$ GeV neutrinos.
\begin{table}[hbt]
\caption{Summary of $R$ measurements, $E_\nu \sim 1$ GeV.}
\label{tab:r-summary}
\begin{center}
\begin{tabular}{lrrl}
\hline
\multicolumn{1}{c}{Experiment} & 
\multicolumn{1}{c}{kt-yr} & 
\multicolumn{1}{c}{events} & 
\multicolumn{1}{c}{R (data/MC)} \\ 
\hline
Super-K\cite{SK98-subgev}
& 25.5 & 1853 & $0.61\pm.03\pm.05$ \\
IMB\cite{Casp91-IMB}
& 7.7 & 610 & $0.54\pm.05\pm.11$ \\
Kam.\cite{Hira88-Kam}
& 7.7 & 482 & $0.60^{+.06}_{-.05}\pm.05$ \\
Soudan-2\cite{Kafk97-taup}
& 3.2 & $\sim$200 & $0.61\pm.15\pm^.05$ \\
Fr\'{e}jus\cite{Daum95-Frejus}
& 2.0 & 200 & $1.00\pm.15\pm.08$ \\
NUSEX\cite{Agli89-NUSEX}
& 0.7 & 50 & $0.96^{+.32}_{-.28}$ \\
\hline
\end{tabular}
\end{center}
\end{table}

%%%\enlargethispage*{.5in}

Kamiokande also studied events with $E_{vis} >$ 1.33 GeV (multi-GeV)
and included partially contained (PC) events where a track was
detected exiting the inner detector\cite{Fuku94-mgev}. They measured a
low value for $R$ of $0.57^{+0.08}_{-0.07}\pm0.07$, but more
interesting was the dependence of $R$ on zenith angle. Neutrinos that
travelled $\sim10^4$ km from below showed a small value of $R$, but
those that travelled $\sim10$ km from above agreed with expectation,
suggesting an oscillation length somewhere in between.

Most of the IMB multi-GeV exposure had a restriction on the
maximum number of PMT hits (to concentrate on proton decay); the
restriction was eventually removed, so they made a separate
analysis\cite{Clar97-IMBmgev} of their last 2.1 kt$\cdot$years for
$E_{vis}>.95$ GeV and measured $R=1.40^{+0.45}_{-0.34}\pm0.14$ with no
zenith dependence. The caveats are: limited statistics of 72 events
(some overlapping with Ref.\cite{Casp91-IMB}); no outer detector to
help identify PC muons; coarser sampling (4\% photon coverage),
resulting in only 90\% correct $e/\mu$ identification.

\section{Soudan-2}

Until recently, atmospheric neutrino results seemed to be divided
between water Cherenkov detectors~\cite{Casp91-IMB,Hira88-Kam}
(anomalous) and iron calorimeters~\cite{Daum95-Frejus,Agli89-NUSEX}
(as expected).  The Soudan-2 collaboration, which operates a
fine-grained iron tracking calorimeter in Minnesota, U.S.A., has
recently published results~\cite{Alli95-soudan} which support the
anomaly seen in the water Cherenkov experiments.

At this meeting, T. Kafka has updated the results from Soudan-2 to 3.2
kt$\cdot$years\cite{Kafk97-taup}. They measure 91 single-prong track
events (mostly charged current (CC) $\nu_\mu$) with $p>100$~MeV/c, and
137 shower events (mostly CC-$\nu_e$) with $p>150$~MeV/c. The
interaction vertex is allowed as close as 20 cm from the edge of the
detector; with only 32 gm/${\rm cm^2}$ of shielding they observe a
significant (25-30\%) background from gamma rays and neutrons
associated with nearby cosmic rays. However, they use an active shield
of proportional tube planes lining the detector hall to veto most
nearby cosmic rays, as well as separately estimate the remaining
background rate as a function of flavor, depth into the detector, and
energy.

After background subtraction, the number of shower events matches
their Monte Carlo prediction; however they observe 37\% fewer tracks
than they predict. Since the total flux is uncertain, it is better to
consider the double-ratio, which they measure to be
$R=0.61\pm0.15\pm0.05$. Regarding the atmospheric
neutrino anomaly, this is a considerable new piece of information, as
the systematics are very different from the water Cherenkov
detectors. Although there are no demonstrated nuclear effects that
would change the ratio of $\nu_\mu$ to $\nu_e$ cross
sections\cite{Enge93-Fermigasmodel}, it interesting that the anomaly
has also been seen in Fe as well as in ${\rm H_2O}$. What the
difference is between these results and those of Fr\'{e}jus and NUSEX
(beyond what may be encompassed by large uncertainties) remains to be
explained.

\section{Super-Kamiokande}

Super-Kamiokande is the next generation water Cherenkov experiment
after IMB and Kamiokande. The detector resides nearby the old
Kamiokande detector in a mine near Kamioka, Japan. However, it is much
larger (22.5 kton fiducial mass, versus 1 kton for Kamiokande and 3.3
kton for IMB). It is instrumented with 11,146 PMTs, each 50 cm across,
such that 40\% of the inner surface area is active photocathode. The
PMTs and electronics are of advanced design, with 2.5 ns RMS timing
for single photoelectrons. The inner detector is surrounded by an
outer volume of water $\sim$2.7 m thick that shields against incoming
radioactivy and is instrumented with PMTs to tag penetrating muons.
Further description of the detector, as well as new measurements of
the $^8B$ solar neutrino flux were presented at this conference by K.
Inoue\cite{Inou97-taup}.

The measurement from Super-Kamiokande is the result of a 414.2
live-day exposure (25.5 kton$\cdot$yrs) during the period from May
1996 to October 1997. The data is reduced from approximately 800K
events per day to about 30 events per day by a series of software
cuts. The most powerful requirement is the absence of hits in the
outer detector, which indicates a fully contained (FC) interaction.
The remaining events are then filtered by a visual scan, where the
principal backgrounds are: (1) cosmic ray muons that evade the outer
detector veto, typically by entering along cable bundles and then
stopping in the detector, and (2) ``flashing'' PMTs that emit light
due to internal corona discharge. The partially contained sample is
formed by a different reduction program from the same
exposure\footnote{The results quoted here are updated from previous
  conference presentations, where PC livetime was somewhat less than
  FC livetime, and PC data was scaled (by $\sim1.1$) when FC+PC
  results were plotted.}, since outer detector hits are now expected,
and the background rejection of entering cosmic rays is different. A
10.0 live-year Monte Carlo sample of $\nu$ interactions is passed
through the same reduction chains, except for the visual scan.

%%% For the
%%%results presented here, the livetime of the PC sample is 370.8
%%%live-days.

The remaining events, both data and Monte Carlo, are passed through
the same reconstruction code to: (a) fit the vertex of the
interaction, by residual PMT timing, (b) count the number of Cherenkov
rings, (c) estimate the direction of each ring, (d) estimate the
energy of each ring, (e) determine the particle type
($\mu$-like,$e$-like) for each ring, and (f) count the number of
$\mu$-decay electrons that follow each event. Most of the analysis is
then done with the sample of events in which the number of rings
found in (b) is exactly one. In most cases, this is the
final state lepton from a charged current neutrino interaction; the
principal contamination is single pion production associated with
neutral current (NC) interactions.

The absolute energy scale was determined to $\sim 2.5\%$ accuracy
using several calibration signatures: LINAC electrons, radioactive
sources, $\pi^0$s and cosmic ray muons. About 9 photoelectrons are
recorded for 1 MeV of visible energy. Conversion to lepton
momentum takes into account the Cherenkov cutoff for muons.

Data samples are defined using the same kinematic criteria as in the
Kamiokande experiment: $p_e>100~\mbox{MeV/c}$, $p_\mu>200~\mbox{MeV/c}$,
and $E_{vis}\le1.33~\mbox{GeV}$ for sub-GeV; $E_{vis}>1.33~\mbox{GeV}$
for multi-GeV FC. The partially contained sample is specified by a
vertex in the inner detector and correlated hits in the outer
detector; the minimum visible energy required is $\sim350$ MeV.
Because $\langle E_\nu \rangle$ is $>>1$ GeV for the PC sample, these
data are added to to the FC multi-GeV sample; the CC lepton is assumed
to be a muon, and no single-ring is required. The fiducial sample is
restricted to events with vertex 2 m from the PMT wall (22.5
kton).

\begin{table*}[hbt]
% space before first and after last column: 1.5pc
% space between columns: 3.0pc (twice the above)
\setlength{\tabcolsep}{1.5pc}
\newlength{\digitwidth} \settowidth{\digitwidth}{\rm 0}
\catcode`?=\active \def?{\kern\digitwidth}
% -----------------------------------------------------
\caption{Event summary for 25.5 kt$\cdot$year sample of
fully-contained atmospheric neutrinos in
Super-Kamiokande. Monte Carlo breakdown uses Honda flux.}
% -----------------------------------------------------
\label{tab:sk-event-summary}
\begin{tabular*}{\textwidth}
{@{}l@{\extracolsep{\fill}}rrr@{\hspace{.2in}}@{\extracolsep{\fill}}rrr}
\hline
 & & \multicolumn{2}{c}{Monte Carlo prediction} 
 & \multicolumn{3}{c}{Monte Carlo breakdown} \\
 & Data & Bartol\cite{Barr89-bartolflux} & Honda\cite{Hond95-flux} 
 & CC-$\nu_\mu$ & CC-$\nu_e$ & NC \\
\hline
sub-GeV $e$-like & 983 & 788.9 & 812.2 & 2\% & 88\% & 10\% \\
sub-GeV $\mu$-like & 900 & 1185.4 & 1218.3 & 96\% & 0.5\% & 4\% \\
sub-GeV multi-ring & 696 & 753.7 & 759.2 & 43\% & 24\% & 33\% \\
\hline
multi-GeV $e$-like & 218 & 190.9 & 182.7 & 7\% & 84\% & 9\% \\
multi-GeV F.C. $\mu$-like & 176 & 229.7 & 229.0 & 99\% & 0.5\% & 0.4\% \\
multi-GeV P.C. ($\mu$-like) & 230 & 305.0 & 287.7 & 98\% & 1.5\% & 0.6\% \\
multi-GeV multi-ring & 398 & 450.1 & 433.6 & 55\% & 30\% & 15\% \\
\hline
\end{tabular*}
\end{table*}

The event totals are listed in Table~\ref{tab:sk-event-summary} along
with the totals for the Monte Carlo samples, scaled to 25.5
kton$\cdot$yrs. These yield the following values of $R =
(N_\mu/N_e)_{DATA}/(N_\mu/N_e)_{MC}$:
\[ \mbox{sub-GeV} \left\{ \begin{array}{rl}
0.610^{+0.029}_{-0.028}\pm0.049 & \mbox{(Honda flux)} \\
0.609^{+0.029}_{-0.028}\pm0.049 & \mbox{(Bartol flux)}
\end{array}
\right. \]
\[ \mbox{multi-GeV} \left\{ \begin{array}{rl}
0.659^{+0.058}_{-0.053}\pm0.081 & \mbox{(Honda flux)} \\
0.665^{+0.059}_{-0.053}\pm0.082 & \mbox{(Bartol flux)}.
\end{array}
\right. \]
For both the high and low energy samples there is a significant
deviation of the $\mu/e$ ratio from the expected value of 1.  The
leading contributions to the systematic uncertainty in $R$ are:
($\nu_\mu/\nu_e$) flux (5\%), neutrino cross section (4.6\% for
sub-GeV and 5.8\% for multi-GeV), and single-ring selection (3\% for
sub-GeV and 6\% for multi-GeV).

It is informative to check the relative rate of $\mu$-decay associated
with the event sample. The decay electrons are detected as time
separated hits from the neutrino interaction; most come either from
associated $\pi^+ \rightarrow \mu^+ \rightarrow e^+$, or directly from
$\mu^{\pm}\rightarrow e^{\pm}$ in CC-$\nu_\mu$ interactions. The
performance of $e/\mu$ identification algorithms is no longer in question
\cite{Kasu96-pid}, but the measured $\mu$-decay fractions check that
the associated pion production is reasonably modeled in the Monte
Carlo. Table~\ref{tab:sk-mudecay} shows that the expected fraction of
$\mu$-decay agrees well with the prediction; the fraction of
$\mu$-decay found in stopping cosmic ray muons verifies the efficiency
of the reconstruction.

\begin{table}[hbt]
\caption{$\mu$-decay fractions.}
\label{tab:sk-mudecay}
\begin{tabular*}{\columnwidth}{lrr}
\hline
\multicolumn{3}{c}{Percentage of events with $\ge1$ $\mu$-decay} \\
\multicolumn{1}{c}{} & 
\multicolumn{1}{c}{Data} & 
\multicolumn{1}{c}{Monte Carlo} \\ \hline
stopping CR $\mu$ & $74.0\pm0.3\%$ & $72.9\pm0.4\%$ \\
$\mu$-like & $67.6\pm1.6\%$ & $68.1\pm0.1\%$ \\
$e$-like & $9.3\pm0.9\%$ & $8.7\pm0.3\%$ \\ \hline
\multicolumn{3}{c}{Percentage of events with $\ge2$ $\mu$-decay} \\
\multicolumn{1}{c}{} & 
\multicolumn{1}{c}{Data} & 
\multicolumn{1}{c}{M.C.} \\ \hline
stopping CR $\mu$ & $0.0\pm0.0\%$ & $0.0\pm0.0\%$ \\
$\mu$-like & $2.9\pm0.6\%$ & $4.1\pm0.1\%$ \\
$e$-like & $0.2\pm0.1\%$ & $0.1\pm0.0\%$ \\
\hline
\end{tabular*}
\end{table}

%%%\enlargethispage*{.5in}

The Super-Kamiokande group had two independent analysis efforts that
were used to check each other and minimize the possibility that some
mistake would be made. The data were separated after electronics
calibration of the PMT data to photoelectrons and nanoseconds.
Otherwise, everything was coded independently, including event
reduction and reconstruction, Monte Carlo generation, and estimation
of the energy scale. Beyond the independent code, the major
differences in the second analysis were: (a) data reduction involved
no scanning, (b) single-ring selection was based on an algorithm that
classified events as single-ring or multiple-ring without attempting
to count the number of rings, (c) $e/\mu$ identification was somewhat
simpler and less efficient (97\% vs $>$99\%), (d) some details of
vertex and direction reconstruction were different.

Upon comparison, the independent analyses were in exceptional
agreement. Of the sub-GeV events found in the fiducial volume by the
second analysis, 99.9\% were found in the data sample of the first
analysis, with 89\% in the fiducial volume, consistent with the vertex
fit resolution. Single-ring classification was in agreement 90\% of
the time. Comparing common events in the fiducial single-ring sub-GeV
sample, vertices agreed to 84 cm RMS, direction agreed to 2.5$^\circ$,
momentum agreed to 0.5\%, 97\% of the events agreed in particle
identification. The value of $R$ for the sub-GeV sample of the second
analysis is: 0.65$\pm$0.03$\pm$0.05; the difference in value from the
first analysis is understood to be due to differences in analysis
methods and within their systematic uncertainties. In sum, the
independent analysis provides reassurance that the deviation of $R$
from unity is not due to experimental mistakes.

To consider that the anomalous $R$ is due to neutrino oscillation,
one looks for a path length or energy dependence of the effect.
The probability of two-flavor neutrino oscillation from $\nu$ to $\nu^\prime$
is given by:
\begin{equation}
P_{\nu\nu^\prime} = \sin^2 2\theta \sin^2 1.27
\frac{\Delta m^2(\mbox{eV}^2/c^2) L(\mbox{km})}{E(\mbox{GeV})},
\end{equation}
where $\theta$ and $\Delta m^2 \equiv |m^2_\nu - m^2_{\nu^\prime}|$
are fundamental parameters that govern the neutrino mixing, and $L$
and $E$ are the path length and energy of the neutrino. The final
state lepton direction and energy are correlated with the incoming
neutrino; for the sub-GeV sample, the mean opening angle for
$\nu_\mu$--$\mu$ is $54^\circ$, for $\nu_e$--$e$ it is $62^\circ$; for
the multi-GeV sample it improves to $<15^\circ$.

\begin{figure}[!tbp]
\centerline{\psfig{figure=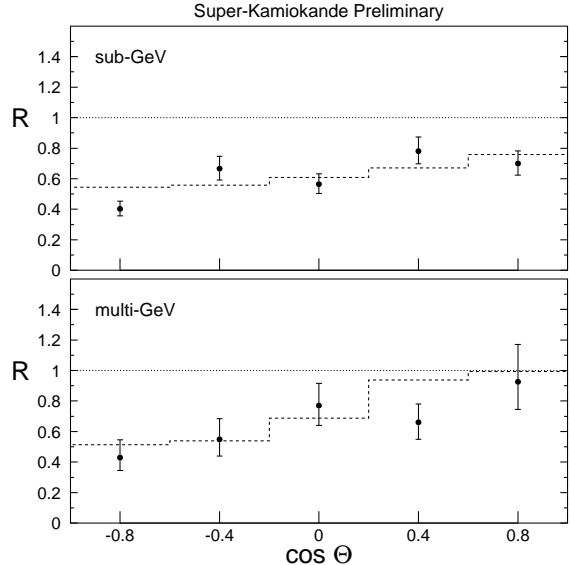,height=2.7in}}
\caption{The zenith angle dependence of $R$ for sub-GeV and
  multi-GeV atmospheric neutrino samples from Super-Kamiokande. The
  dashed line shows the expected shape for $\nu_\mu \rightarrow
  \nu_\tau$ oscillation with $\sin^2 2\theta = 1$ and $\Delta m^2 =
  .005$ ${\rm eV^2/c^2}$.}
\label{fig:r-zen}
\end{figure}

The samples are divided into 5 $\cos \Theta$ bins where $\Theta$ is
the angle between the outgoing lepton direction and the
nadir\footnote{Caveat: the IMB collaboration used the opposite
definition, so down-going neutrinos are near $\cos \Theta = -1$ in
their publications.}; so down-going neutrinos that are produced
directly overhead, with short travel distance, populate the bin near
$\cos \Theta = 1$. Calculating $R$ for each zenith bin results in
Fig.~\ref{fig:r-zen}. A slight asymmetry is evident in the sub-GeV
sample, and a strong asymmetry is evident in the multi-GeV sample. If
there were no anomaly, the $R$ values would be around 1; for
hypothetical oscillation parameters of $\sin^2 2\theta = 1$ and
$\Delta m^2 = .005$ ${\rm eV^2/c^2}$, the dashed line is expected, and
is a better match to the data.

\begin{figure}[!tbp]
\centerline{\psfig{figure=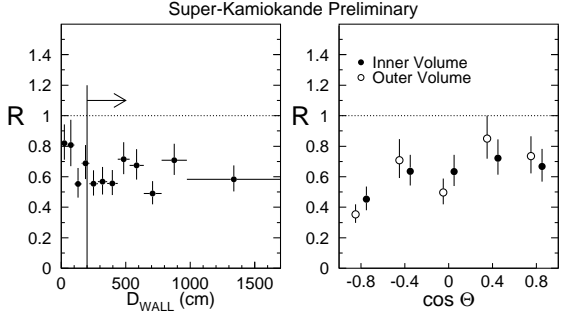,height=1.35in}}
\caption{(A) $R$ for the sub-GeV samples as a function of distance
from the PMT wall ($D_{WALL}$). (B) $R$ versus zenith angle for two
concentric fiducial volumes: $5 > D_{WALL} > 2$ meters (outer) and
$D_{WALL} > 5$ meters (inner).}
\label{fig:r-fid}
\end{figure}

It is interesting to check the result as a function of position in the
detector because (a) some reconstruction algorithms are less
certain for vertices close to the PMT wall, and (b) possible
backgrounds, such as neutrons from nearby cosmic rays
\cite{Ryaz95-neutron,Fuku96-neutron}, or
entering events that evade the outer detector veto, would accumulate
near the fiducial boundary.  Figure~\ref{fig:r-fid}a shows $R$ versus
distance from the PMT wall, where the fiducial volume is found at 2
meters.  Figure~\ref{fig:r-fid}b shows the zenith angle dependence of
R, dividing the data into two approximately equal fiducial volumes: an
outer volume between 2--5 m from the PMT wall and an inner volume
greater than 5 m from the wall. There is no variation of the result
due to the fiducial boundary evident in either figure.

\begin{figure}[!tb]
\centerline{\psfig{figure=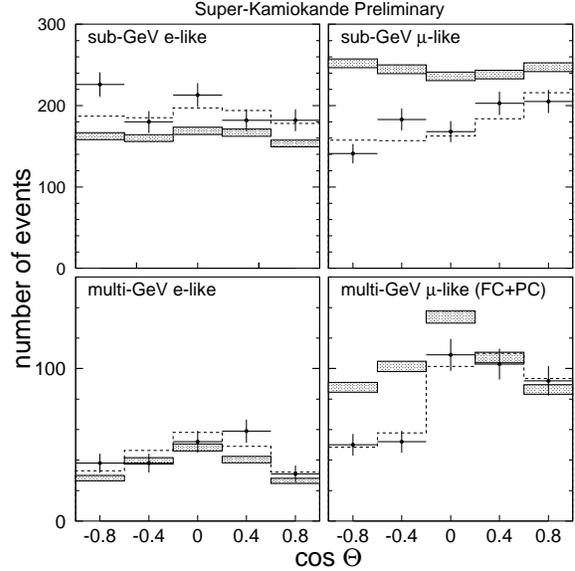,height=2.7in}}
\caption{The number of $\mu$-like and $e$-like events
  as a function of zenith angle. The solid histograms are the Monte
  Carlo expectation with no neutrino oscillation; the thickness
  represents the statistical uncertainty in the Monte Carlo sample.
  The dashed line shows the expected shape for $\nu_\mu \rightarrow
  \nu_\tau$ oscillation with $\sin^2 2\theta = 1$ and $\Delta m^2 =
  .005\mbox{eV}^2/c^2$.}
\label{fig:rate-zen}
\end{figure}

The double-ratio of $(\mu/e)_{DATA}$ to $(\mu/e)_{MC}$ is useful to
illustrate the effect, but it does not indicate whether $\mu$ or $e$
rates (or both) are affected. Furthermore, $R$ is not so practical for
statistical tests. Figure~\ref{fig:rate-zen} shows the $\mu$ and $e$
rates separately for sub-GeV and multi-GeV (FC+PC) compared to Monte
Carlo prediction. The solid bands are the absolute prediction, where
the height of the band is equal to the Monte Carlo statistical
uncertainty. Not shown is the $\pm20\%$ normalization uncertainty,
which is highly correlated bin-to-bin, between $\mu$ and $e$, and
between sub-GeV and multi-GeV. Even accounting for this uncertainty,
it is apparent that the anomalous $R$ is dominated by a deficit of
$\mu$-like events coming from below ($\cos \Theta <0$).

\enlargethispage*{0cm}

The significance of this result can be easily evaluated by calculating
the up-down asymmetry $A = (N_{down}-N_{up})/(N_{down}+N_{up}$) where
up and down are defined by $\cos \Theta$ $<-.2$ and $>.2$
respectively (Tab.\ref{tab:asym}). Besides other interesting
possibilities\cite{Flan97-asym}, the distribution of $A$ is nicely
described by a gaussian variance. For multi-GeV events, 
$N_{up}$ and $N_{down}$ should be nearly
symmetrical, whereas $A$ for $\mu$-like data (FC+PC) differs from
expectation by greater than $5\sigma$.
\begin{table}[hbt]
\caption{Up-down asymmetry for multi-GeV data.}
\label{tab:asym}
\begin{tabular*}{\columnwidth}{lrrr}
\hline
\multicolumn{1}{c}
{Sample} & $N_{up}$ & $N_{down}$ & \multicolumn{1}{c}{$A$} \\ 
\hline
$\mu$-like data & 102 & 195 & 0.313$\pm$0.055 \\
$e$-like data & 76 & 90 & 0.084$\pm$0.077 \\
$\mu$-like MC & 1669 & 1707 & 0.013$\pm$0.017 \\
$e$-like MC & 596 & 589 & -0.006$\pm$0.029 \\
\hline
\end{tabular*}
\end{table}

The dashed line in Fig.\ref{fig:rate-zen} represents an
oscillation hypothesis\footnote{The values $(1,.005)$ represent a
  test point; the exact best fit location can change when the
  technique or data sample changes since the minimum is fairly flat.}
of ($\sin^2 2\theta=1$,$\Delta m^2=.005$) for $\nu_\mu$ disappearance.
The overall normalization is adjusted upward (thus the $e$-like rate
increases to better match the data, even when $\nu_e$ mixing is not
considered) while (2) is used to calculate the probability of
$\nu_\mu$ disappearance and reweight the Monte Carlo. The $\nu$ travel
distance $L$ is calculated as a function of energy and flavor based on
a production height model\cite{Gais97-prodhgt}. The oscillation
hypothesis provides a reasonable fit to the data, certainly better
than the null hypothesis.

\enlargethispage*{0cm}

\begin{figure}[!tb]
\centerline{\psfig{figure=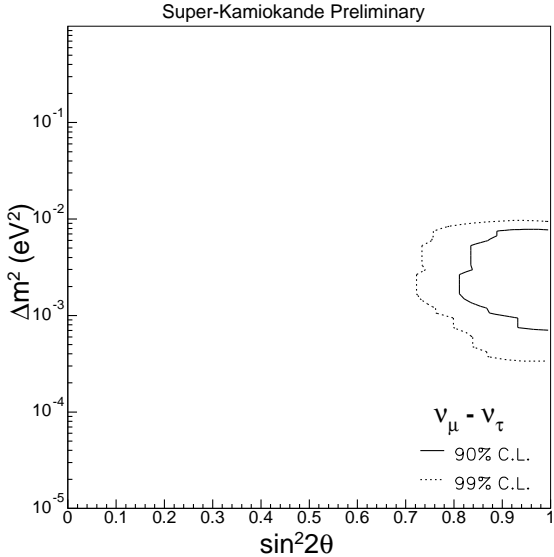,height=2.5in}}
\caption{Confidence intervals for $\sin^2 2\theta$ and $\Delta m^2$
based on a $\chi^2$ fit to Super-Kamiokande atmospheric
neutrino data binned by lepton identification, lepton momentum
and $\cos \Theta$. The solid line is 90\% CL, the dashed line is 99\% CL.}
\label{fig:sk-contour}
\end{figure}

The exact details of fitting the data to estimate possible mixing
parameters are still being evaluated. Using a method similar to that
used by Kamiokande\cite{Fuku94-mgev}, $\chi^2$ terms are formed
between the data and Monte Carlo prediction [modified by
$P_{\nu\nu^\prime}(\sin^2 2\theta,\Delta m^2)$],
binned in zenith angle, energy, and flavor (values of $R$ are not used
directly in the fit). The normalization, $N_\mu/N_e$ ratio and
systematic terms are allowed to adjust and contribute to the
$\chi^2$. For the Super-Kamiokande data, the minimum $\chi^2$ is found
to be rather likely, $\sim$30\% depending on the details.
From the fit, a confidence interval is drawn based on
$\chi^2_{min}+4.6$ (90\% CL) such as shown in Fig.~\ref{fig:sk-contour}.

\enlargethispage*{0cm}

The preferred interval for $\sin^2 2\theta$ is found near $1$ because
the upward zenith bin with $\langle L\rangle \sim 10,000$ km,
presumably has averaged over several oscillation lengths and
$P_{\nu\nu^\prime}$ is one half. In a scenario where more than two
$\nu$ flavors are mixing, the average value can be less than one half.
The $\Delta m^2$ range is determined by the shape of the zenith angle,
also considering dependence on $E_\nu$. This 90\% confidence interval
from Super-Kamiokande prefers a lower range in $\Delta m^2$ than that
found by the Kamiokande collaboration\cite{Fuku94-mgev}, which had a
minimum $\Delta m^2$ of $.005$.

\section{Upward-Going Muons}

The above discussion covered neutrino interactions in the fiducial
volume of the detector. The other class of atmospheric neutrino event
studied is that of $\nu_\mu$ interactions in the rock around the
detector, where the final state muon enters the sensitive region of
the detector. To separate these from ordinary cosmic ray muons, the
muon must be upward-going, or come from the direction of a known
thick overburden. The parent neutrino energy is $10$--$1000$ GeV,
significantly higher than for contained events.

There are several current measurements of the total rate of
upward-going muons. In addition to the water Cherenkov detectors
described elsewhere in this paper, MACRO and Baksan are large area
scintillator detectors that distinguish upward-going muons by
time-of-flight; T. Montaruli (MACRO) and S. Mikheyev (Baksan) have
presented updated results at this conference. The measured and
expected event rates are compared in Table~\ref{tab:upmu}; the
experiments are listed in order of increasing absorber depth for
directly vertical muons. The uncertainty in the measured number of
events is statistical only; some experiments have estimated that the
uncertainty due to experimental systematics could be as large as
8\%. The uncertainty in the expected number of events is a common
15--20\%, due mostly to the uncertainty in the absolute flux.
Considering this, in no case is a significant deficit of muons
measured, but the measurements are generally low compared to
expectation.
\begin{table}[hbt]
\caption{Summary of through-going upward-$\mu$ totals.}
\label{tab:upmu}
\begin{tabular*}{\columnwidth}{lrr}
\hline
\multicolumn{1}{c}{} & 
\multicolumn{2}{c}{Number of Events} \\ 
\multicolumn{1}{c}{Experiment} & 
\multicolumn{1}{c}{Measured} & 
\multicolumn{1}{c}{Expected} \\ 
\hline
MACRO\cite{Mont97-taup} & $350\pm19$ & 472 \\
Baksan\cite{Mikh97-taup} & $558\pm24$ & 557 \\
Kamiokande\cite{Kam98-unpub} & $373\pm19$ & 414 \\
IMB\cite{Beck95-nu94} & $539\pm23$ & 550 \mbox{or} 625 \\
Super-K\cite{SK98-unpub} & $410\pm20$ & 445 \\
\hline
\end{tabular*}
\end{table}

\enlargethispage*{0cm}

There are two other approaches that probe neutrino
oscillations\cite{Lipa97-upstopzen} using upward-going muons. IMB
measured the ratio of stopping upward muons ($\langle E_\nu\rangle
\sim 10$ GeV) to through-going upward muons ($\langle E_\nu\rangle
\sim 100$ GeV) to be $0.16\pm.02$. Although much of the flux
uncertainty cancels out, the usual analytic integration using deep
inelastic scattering and parton distribution functions must be handled
with care\cite{Beck95-nu94,Frati93-upmu}, especially at low
energy\cite{Lipa95-upstop}. After these considerations, the predicted
rate was .14 or .18 depending on the flux
model\cite{Beck95-nu94}. Based on the agreement of data with
prediction, a small excluded region in $\sin^2 2\theta$, $\Delta m^2$
was drawn around $\Delta m^2 =10^{-3}\mbox{eV}^2/c^2$, where the
strongest deviation would have been found; this happens
to be in conflict with the region favored by the Super-K
contained vertex data. Super-Kamiokande will also measure the stopping
ratio, with the advantage of a very thick detector that stops a large
number of upward-going muons; the analysis is in progress.

\begin{figure}[!tbp]
\centerline{\psfig{figure=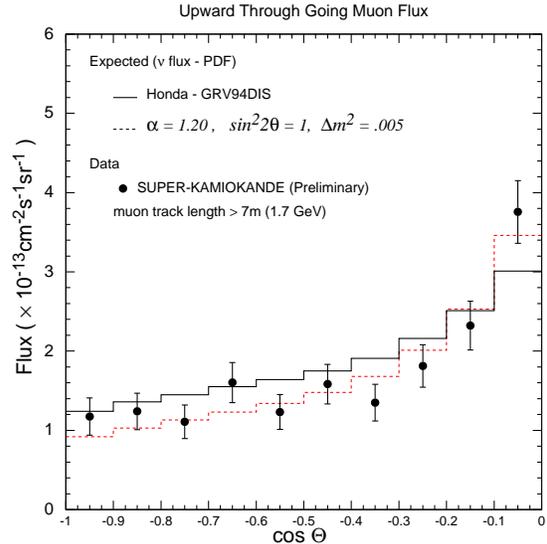,height=2.4in}}
\caption{The through-going upward-going $\mu$ flux as a function of
zenith angle as measured by Super-Kamiokande. The data points are
compared against expectation (solid histogram) and the same
expectation with normalization $\alpha=1.2$ and $\nu$-mixing
$\sin^2 2\theta=1$, $\Delta m^2=.005$.}
\label{fig:sk-upmu}
\end{figure}

The second approach uses the shape of the zenith distribution, which
may be distorted as the baseline varies from 500 km at the horizon to
12,000 km at the nadir. Because the energy spectrum of the parent
neutrinos is broad, $\sim10-1000$ GeV, the change in shape is gradual,
with some steepening at the horizon as the probability decreases for
high energy $\nu_\mu$ to oscillate.  Figure~\ref{fig:sk-upmu} shows
the Super-K measurement of the flux versus zenith angle
compared to expectation; recall that the normalization of the
prediction is uncertain to $\pm$20\%. When the normalization of the
Monte Carlo is decreased by a factor of $\alpha=0.83$ to match the
data, the $\chi^2$ is 12.7; alternatively, when the normalization is
increased by a factor of $\alpha=1.2$, and $\nu_\mu$ disappearance
oscillations are applied with $\sin^2 2\theta=1$, $\Delta m^2=.005$,
the fit is somewhat better, with a $\chi^2$ of 8.3.

The other experiments listed in Table~\ref{tab:upmu} can make this
comparison as well.  The scintillator detectors MACRO and Baksan
unfortunately have reduced acceptance at the horizontal, so those bins
require significant geometric correction. Curiously, even the
well-measured upward bins of those two experiments are not smooth and
suffer from poor agreement in shape with: (a) no oscillation, (b) any
2-flavor oscillation parameters\footnote{A recent preprint considers
  that matter oscillation with a sterile neutrino crossing the Earth's
  core may modulate the prediction with features similar to the MACRO
  data\cite{Liu97-mswupmu}.}, and (c) each other. Both experimental
groups have made extensive checks for a systematic error, but have
found no cause\cite{Mikh97-taup,MACR98-unpub}.

\section{Conclusions}

Currently, the evidence for an anomalous ratio of atmospheric neutrino
flavors is inconsistent across experiments. Of course, prior results
remain intact, and either support or disagree with the anomaly. But
the latest results support that a significant discrepancy exists
between experiment and prediction.  The measurement by Soudan-2 shows
that the anomaly is not specific to water Cherenkov detectors. A
zenith angle measurement of the Soudan-2 data could be very
interesting.  Significant new information is taken from the high
statistics Super-Kamiokande data: the anomaly is strongly confirmed in
$R$ and the zenith angle dependence of $R$. The shape of the zenith
angle dependence is very suggestive of neutrino oscillations. The
multi-GeV $\mu$-like rate as a function of zenith angle
indicates that $\nu_\mu$ disappearance is favored over $\nu_\mu
\rightarrow \nu_e$ oscillation.  Even though 1.1 years of
Super-Kamiokande running has surpassed the prior world statistics,
more livetime will be welcome and allow finer subdivision of the data
for cross-checks and estimation of possible mixing parameters.

Still, to claim the atmospheric $\nu$ anomaly is caused by neutrino
mixing requires confirmation with as many other signatures as
possible. Upward-going muons currently show a mix of results, none
obviously refuting or confirming. The upward $\mu$ flux measurement of
Super-Kamiokande is consistent with the estimated parameters from the
contained vertex sample, but the result is not conclusive. Further
analysis of stopping muons and the angular distribution is awaited.

In the end, the best confirmation should come from long-baseline
neutrino beams from accelerators. However, according to the
preliminary Super-K results, the experiments under
construction may have to address lower values of $\Delta m^2$ than
when they were conceived.

\section{Acknowledgements}
The author thanks the conference organizers and the participants
listed in this paper. He is especially appreciative of his collaborators
on Super-Kamiokande for preparing the latest results, which are presented
here on their behalf. He gratefully acknowledges financial support by the
U.S. Department of Energy.

\end{document}